# Ionization and excitation of low-lying circular states of the hydrogen atom in strong circularly polarized laser fields

Jarosław H. Bauer[1,*] and Zbigniew Walczak[1]

[1]*Department of Theoretical Physics, Faculty of Physics and Applied Informatics, University of Lodz, Pomorska 149/153, 90-236 Lodz, Poland*

We perform *ab initio* calculations for the hydrogen atom initially in one of the six circular bound states with the principal quantum numbers $n = 2, 3$ and 4, irradiated by a short circularly polarized laser pulse of $400$ $nm$. The field propagates in the direction parallel to the $z$-component of the angular momentum of the atom. For peak intensities in the range of $\sim 10^{11} - 10^{16} W/cm^2$, we investigate probabilities for the atom to ionize or to get on some bound (excited) state or to remain in the initial state after the end of the laser pulse. In most cases, we find pronounced differences in ionization probabilities for atoms in states having different signs of magnetic quantum number. Usually electrons corotating (with respect to the laser field) ionize faster than their counter-rotating equivalents. This is usually unlike in tunneling domain, where counter-rotating electrons always ionize faster. Adiabatic or nonadiabatic tunneling ionization is absent in our case. Near $10^{16} W/cm^2$ these differences in ionization probabilities drop away. We have found important difference in the behavior of the excitation (as a function of the peak laser intensity) for initial states with $n = 2$ and $n \neq 2$ (3 or 4). For higher $n$ the excitation is always weaker than the ionization and starts for higher intensities. For $n = 2$ the strong excitation appears before strong ionization due to large probability of one-photon absorption combined with population of many bound states having principal quantum numbers $n$ from 3 up to several dozen. Quite accurate analysis of the excitation process is presented.

___________________________

*jaroslaw.bauer@uni.lodz.pl

# I. INTRODUCTION

Circular states of an atom are Rydberg states with the highest possible absolute value of the magnetic quantum number $m$ of the bound state described by three well-known quantum numbers $(n,l,m)$ (in a spherical coordinate system). Customarily, $n$ is the principal quantum number, and $l$ is the orbital quantum number. For a given $n$ there are always two circular states, namely $(n,l=n-1,m=n-1)$ and $(n,l=n-1,m=-(n-1))$. For these two states an electron has the maximum absolute value of the $z$-component of angular momentum $|m|\hbar = (n-1)\hbar$. Such excited states of the hydrogen atom can be created by adiabatic transfer of a Rydberg state in crossed electric and magnetic fields or by adiabatic passage in a rotating microwave field [1] (see also references therein for a description of different methods). Circular states can be produced also by utilizing adiabatic switching of orthogonal electric and magnetic fields [2]. These states have long radiative lifetimes because there is only one channel to which they can decay, namely the next lower circular state. Circular states have been produced in most of the alkali atoms and in the hydrogen atom [1]. When $n \gg 1$ behavior of the excited electron may be described approximately by classical concepts, for example, when atom interacts with circularly polarized (CP) microwave radiation [3]. However, the lower the principal quantum number $n$ is, the stronger becomes the demand that the atom should be fully described by quantum mechanics, namely by the Schrödinger equation.

In this work, we investigate a behavior of the hydrogen atom, initially in a low-lying circular state ($2 \le n \le 4$), interacting with the pulse of the CP electromagnetic radiation. We use nonrelativistic theory and the dipole approximation to describe the laser field. Our purpose is to solve numerically (exactly) the time-dependent Schrödinger equation (TDSE) within these approximations. The problem that we study is limited to the situation when the quantization axis of the hydrogen atom (the $z$ axis) is parallel to the propagation direction of the laser field. Both the electron (in any circular state that we consider) and any incoming photon carry angular momenta which are mutually parallel or anti-parallel. In other words, the laser field is polarized in the $xy$-plane. The atom may be ionized, excited or may remain in its initial state after termination of the radiation pulse. Typically, for most atoms (if they are in a ground state) and visible radiation, there are three adjacent regimes of the radiation intensity. From the point of view of a dominant physical mechanism one has to deal with (i) multiphoton ionization (MPI) (for weak laser fields), (ii) tunneling ionization (TI) (for intermediate laser fields) and finally (iii) barrier-suppression ionization (BSI), called also over-the-barrier ionization, (for strong and superstrong laser fields) [4]. For sufficiently intense fields and higher frequencies, one can also speak about above threshold ionization (ATI) of atoms [5]. In experiments made about forty years ago [6-8] one observed additional peaks (separated by the photon energy) in the energy spectrum of outgoing photoelectrons. In the present work, we use atomic units ($a.u.$): $\hbar = e = m_e = 1$, substituting explicitly $-1$ for the electronic charge.

However, for the laser frequency that we have chosen ($\omega = 0.114$ $a.u.$) and low-lying circular states of the hydrogen atom, there are only two regimes of radiation intensity, namely (i) MPI and (iii) BSI. The absence of (ii) TI is connected with the fact that for $n \geq 2$ the binding energy of the atom is very small, hence only very low frequency radiation (in infra-red domain) enables TI. In the paper [9], which preceeds the present one, the following criterion was identified:

$$\omega > \omega_{\lim} \equiv \frac{F_{BSI}}{\sqrt{2E_B}} = \frac{1}{Z}\sqrt{\frac{E_B^3}{2^5}} = \frac{Z^2}{2^4 n^3} \; , \tag{1}$$

where $F_{BSI} = E_B^2/(4Z) = Z^3/(16n^4)$ is the barrier-suppression field strength (see, for example, [5,10,11]), and $E_B = Z^2/(2n^2)$ is the binding energy of the atom (of the nuclear charge $Z$; in this work $Z = 1$). When Eq. (1) is satisfied, TI is negligible but there is even no nonadiabatic tunneling regime [12,13]. TI was first described by Keldysh [14,15], who identified the parameter (termed after him) $\gamma = \omega\sqrt{2E_B}/F = Z\omega/(nF)$ ($F$ is an amplitude of the laser field). Roughly up to $\gamma < 1$ (usually $\gamma \ll 1$) TI or nonadiabatic tunneling ($\gamma \sim 1$) is possible, but only if $\omega$ is sufficiently low. This is not the case considered in this work, where Eq. (1) is very well satisfied.

We will call the laser light 'corotating with respect to the initial state' of the hydrogen atom if the electron and the CP photon angular momenta are parallel. When these two angular momenta are anti-parallel, we will call the laser light 'counter-rotating with respect to the initial state'. Of course, these two physical situations are different and should lead to different observable effects like, for example, ionization rates, ionization probabilities or photoelectron energy spectra. Indeed, in the case of single photon ionization, corotating light stronger ionizes the atom than counter-rotating light [12,16]. The same concerns Rydberg electrons, although in this case the mechanism (successive excitations of Rydberg states) is different [3,12]. In the nonadiabatic tunneling regime counter-rotating light stronger ionizes the atom than corotating light [12], so the situation is reverse. When Eq. (1) is satisfied and at least three photons are needed to ionize the hydrogen atom this situation depends on the laser intensity. For lower intensities (for example, in perturbative regime) corotating light stronger ionizes the atom, but for higher intensities the ionization by counter-rotating light is stronger (cf. Fig. 5 in Ref. [9]). In this work, we compare ionization probabilities (after termination of the laser pulse) for pairs of low-lying circular states ($2 \leq n \leq 4$) having magnetic quantum numbers $m = -(n-1)$ and $m = (n-1)$, respectively. We also study excitation probabilities (as in Ref. [9]). Both excitation and ionization probabilities are investigated as functions of the peak laser intensity, starting from $10^{11} W/cm^2$ and ending near $10^{16} W/cm^2$. The frequency used in the present work is two times bigger than that used in Ref. [9] ($\omega = 0.057$ $a.u.$). Therefore, we are able to investigate the effect of higher frequencies for the initial states $(2,1,-1)$ and $(2,1,1)$ and the inequality (1) is better satisfied for low-lying circular states.

Our paper is organized as follows. A short exposition of the theory is given in Sec. II. In Sec. III we present and discuss our numerical results, most extensively with respect to the excitation of the atom. We broadly refer to some earlier works utilizing possible physical interpretations of the obtained results. The summary and main conclusions are given in Sec. IV.

## II. THEORY

In the present work we perform accurate *ab initio* calculations (see, for example, Ref. [17] for CP fields). The following TDSE is solved numerically ($\vec{p} = -i\vec{\nabla}$)

$$i\frac{\partial \Psi(\vec{r},t)}{\partial t} = \left[\frac{1}{2}\left(\vec{p} + \frac{1}{c}\vec{A}(t)\right)^2 - \frac{Z}{r}\right]\Psi(\vec{r},t) \qquad (2)$$

with

$$\vec{A}(t) = A_0 f(t)[-\hat{x}\sin(\omega t) + \hat{y}\cos(\omega t)], \qquad (3)$$

where $A_0 = F_0/\omega$ is the amplitude of the vector potential, $\hat{x}$ and $\hat{y}$ are versors in the $x$ and $y$ directions, respectively, and $f(t)$ is a slowly varying pulse envelope which has a sine-squared form:

$$f(t) = \sin^2\left(\frac{\pi t}{t_d}\right), \qquad (4)$$

and the pulse time duration is $t_d$. Let us note that $\vec{A}(0) = \vec{A}(t_d) = \vec{0}$, hence our results are gauge invariant [18] (because only $\Psi(\vec{r}, t_d)$ is utilized for a calculation of various probabilities). The method of solution of Eq. (2) is the same as in Ref. [9] and very similar to the one described in Refs. [19,20]. There are benchmark results for the hydrogen atom in CP laser fields in Refs. [19,20]. We use the velocity gauge (advantages of which are well known [21]) and the dipole approximation, as indicated by Eqs. (2) and (3). The vector potential $\vec{A}(t)$ from Eq. (3) corresponds to $\sigma^+$ polarization, if $\omega > 0$. We use a spectral method to solve the TDSE expanding the total wave function as follows:

$$\Psi(\vec{r}, t) = \sum_{l=0}^{l_{max}} \sum_{m=-l}^{l} \sum_{n=l+1}^{l+1+N} a_{nlm}(t) \frac{1}{r} S_{n,l}^{\kappa}(r) Y_{l,m}(\theta, \varphi). \qquad (5)$$

In this expression, $a_{nlm}(t)$ is a time-dependent coefficient, $S_{n,l}^{\kappa}(r)$ is a Coulomb-Sturmian function of the electron distance to a nucleus $r$, and $Y_{l,m}(\theta, \varphi)$ is the spherical harmonic of the electron angular coordinates $\theta$ and $\varphi$. The functions $S_{n,l}^{\kappa}(r)$ [20] form a discrete and complete set (for any given $\kappa$) in the space of the $L^2$-integrable functions. $N$ is a number of Coulomb-Sturmian functions per $(l, m)$ pair ($N = 300$ is sufficient for all laser field parameters in this work, but smaller $N$ would be sufficient in perturbative regime). In general, $\kappa$ is a real number such that $0 < \kappa < 1$. The proper value of this parameter (as used in this work) should be chosen as $\kappa = 1/n$ (the reciprocal of principal quantum number of the initial state; however, our final numerical results do not depend on $\kappa$ in some neighborhood of $\kappa = 1/n$). We refer the reader for more detail about this method (including its further development) to Refs. [9,20,22]. The method has been generalized to the case of helium in [23]. The ionization and excitation probabilities are calculated by projecting onto bound and continuum states of the hydrogen atom, after switching off the laser pulse. We would like to stress that atomic and laser field parameters are quite similar to those applied in Ref. [9]. As a result, we could utilize some experience gained by one of us (J.H.B) and obtain fully convergent and exact numerical results.

## III. RESULTS AND DISCUSSION
### A. Ionization, excitation or remaining in the initial state

There are probabilities for ionization, excitation, and remaining in the initial state of the hydrogen atom as a function of the peak laser intensity in Fig. 1. In Figs. 1(a)-1(f) we show these probabilities for the initial states $(2,1,-1)$, $(2,1,1)$, $(3,2,-2)$, $(3,2,2)$, $(4,3,-3)$, and $(4,3,3)$, respectively. For the states with $m < 0$ the laser light is counter-rotating with respect to the initial state, while for the states with $m > 0$ the laser light is corotating. All the probabilities are calculated after switching off the laser pulse. The total duration of the pulse (with a sine-squared envelope) is 10 cycles ($\tau = 10 \times 2\pi/\omega$, with $\omega = 0.114$ $a.u.$). The laser wavelength conforms with a second harmonics of the $Ti:sapphire$ laser ($\lambda = 400$ $nm$). We have checked that carrier envelope phase effects are negligible for these pulse parameters. The condition from Eq. (1) is satisfied in our case (the right-hand side is equal to $0.0078$ $a.u.$ for $n = 2$ and even less for $n = 3$ and $n = 4$). In the lowest order of perturbation theory two photons are needed to overcome the ionization threshold at the binding energy $|E_2| = 0.125$ $a.u.$ for the initial states $(n=2,1,\pm 1)$. Only one photon is needed (in the lowest order of perturbation theory) to overcome the ionization threshold at $|E_3| = 0.05556$ $a.u.$ for the initial states $(n=3,2,\pm 2)$. The same concerns the ionization threshold at $|E_4| = 0.03125$ $a.u.$ for the initial states $(n=4,3,\pm 3)$. The excitation is calculated here as a sum of populations over all bound states except the initial one (populations of bound states with $n$ lower than initially are always much smaller than the total population of bound states in our numerical calculations).

Figures 1(a) and 1(b) explicitly show that when the peak intensity of the laser field grows (beginning from the perturbative regime) the excitation grows initially faster than the ionization. This is very similar to what was observed for two times lower laser frequency ($\omega = 0.057$ a.u.) and the same pulse in Ref. [9]. For $\omega = 0.114$ a.u. substantial ionization starts at the peak intensity close to the value of $I_{BSI} = 2F_{BSI}^2 = 1.1 \times 10^{12} W/cm^2$, but the excitation starts at the peak intensity about 10 times smaller. Then both ionization and excitation grow with intensity. For the state $(2,1,-1)$ the excitation peaks near $4 \times 10^{13} W/cm^2$, and for the state $(2,1,1)$ near $8 \times 10^{12} W/cm^2$. In both cases, at these peaks, more than 70% of atoms remain bound but excited after the end of the laser pulse. The rest of atoms is mainly ionized, very little atoms remain in their initial state (the initial-state population is roughly negligible for $I > 10^{14} W/cm^2$). When the intensity exceeds about $10^{15} W/cm^2$ the excitation and the ionization are nearly constant with some $10-20\%$ of the excitation. This is again quite similar to the case of $\omega = 0.057$ a.u. [9]. Figures 1(a) and 1(b) in the present work qualitatively resemble Figs. 1 and 2 from Ref. [9]. Increasing the laser frequency twice leads to increasing the intensity at which the peak of excitation exists (about 7 times for the state $(2,1,-1)$ and about 4 times for the state $(2,1,1)$). However, an overall picture of the behavior of ionization and excitation probabilities as a function of the peak laser intensity remains the same for $\omega = 0.057$ a.u. and $\omega = 0.114$ a.u.

In Figs. 1(c) and 1(d) we show analogous probabilities for the initial states $(3,2,-2)$ and $(3,2,2)$, and in Figs. 1(e) and 1(f) for the initial states $(4,3,-3)$ and $(4,3,3)$, respectively. For the states $(3,2,\pm 2)$ substantial ionization starts at intensities close to analogous intensities for the states $(2,1,\pm 1)$, respectively. But the excitation is much weaker for the states with $n=3$ and $n=4$ (see Figs. 1(c)-1(f)) and starts for much higher intensities, usually well above $10^{15} W/cm^2$. The ionization for the states with $n=4$ starts for intensities higher than the ionization for analogous states with $n=3$ or $n=2$. The excitation does not exceed 30% for the states $(3,2,\pm 2)$ and 40% for the states $(4,3,\pm 3)$. Moreover, unlike for the states with $n=2$, the excitation appears after the ionization (when increasing the intensity) for the states with $n=3$ and $n=4$ (cf. Figs. 1(c)-1(f)). It appears that the value of the BSI intensity is not relevant to the onset of ionization when $n=3$ or $n=4$. For the states $(3,2,\pm 2)$ and $(4,3,\pm 3)$ $I_{BSI} < 5 \times 10^{10} W/cm^2$. The highest intensity shown in Figs. 1(e) and 1(f) is $4 \times 10^{16} W/cm^2$. It was not possible to achieve convergent results for even more intense fields using our present computational resources. However, let us note that the initial-state probability is nearly zero already for $I > 10^{16} W/cm^2$. Perhaps we should not expect very much changes in ionization and excitation for $I > 4 \times 10^{16} W/cm^2$ (cf. Figs. 1(a)-1(d)) and maybe we should rather expect that ionization prevails over excitation for very strong fields.

In Fig. 2 we present ionization probabilities as a function of the peak laser intensity for six different initial states of the hydrogen atom. The states and the numerical data are exactly the same as in Figs. 1(a)-1(f). The purpose of making this plot is a comparison of initial states from the point of view of their inclination to ionization with increasing the peak intensity. From Fig. 2 one can conclude that for perturbative laser fields the fastest ionization is for the states $(3,2,2)$, $(2,1,1)$, $(2,1,-1)$ and $(3,2,-2)$, respectively. The slowest ionization, in perturbative regime, is for the states $(4,3,-3)$ and $(4,3,3)$, respectively. The states with $m > 0$ (corotating light) ionize faster than the analogous states with $m < 0$ (counter-rotating light). This usually holds also for higher intensities. Above $10^{14} W/cm^2$ the state $(3,2,-2)$ has greater ionization rate than the state $(2,1,-1)$ but the two curves intersect again near $3 \times 10^{15} W/cm^2$. Accordingly, the ratio of their ionization rates must be higher or lower than 1, depending on the intensity. Around $7 \times 10^{14} W/cm^2$ the state $(3,2,-2)$ has the highest (among the six ones) ionization probability. When the intensity achieves the greatest values in Fig. 2 ionization probabilities for the states $(n,l,m)$ and $(n,l,-m)$ become nearly identical.

In Fig. 3 we present initial-state probabilities as a function of the peak laser intensity for six different initial states of the hydrogen atom. The states and the numerical data are exactly the same as in Figs. 1(a)-1(f). In this plot we compare initial states from the point of view of their (usually) increasing depletion with increasing the peak intensity. From Fig. 3 one can conclude that the fastest depletion is for the states $(2,1,1)$, $(2,1,-1)$, $(3,2,2)$ and $(3,2,-2)$, respectively. The slowest depletion occurs for the states $(4,3,-3)$ and $(4,3,3)$, respectively. Roughly for $I > 10^{16} W/cm^2$ the initial-state probability after the end of the pulse is very close to zero for each of the six states. The states with $m > 0$ (corotating light) deplete faster than the analogous states with $m < 0$ (counter-rotating light). From the point of view of the order the six curves in Fig. 3 (from left to right and from $(2,1,1)$ to $(4,3,-3)$) the principal quantum number $n$ is the most important and then the sign of the magnetic quantum number $m$ decides. This order is a little different from the similar order of states in Fig. 2. Of course, this is the excitation that makes the difference.

## B. Remarks on theoretical models

To the best of our knowledge, currently there is no simple analytical model that could explain the excitation and ionization dependence on the laser intensity for all atomic and laser field parameters considered in this work. Most of the intensity range presented in Figs. 1-3 (roughly $10^{11} - 10^{16} W/cm^2$) lies in the BSI regime. Both in Ref. [9] and in Ref. [24] (which may be treated as a supplement to [9]) it was found that the ionization mechanism is multiphoton absorption. In Ref. [24] the ionization and excitation process was studied in temporal, angular and energy resolution. In the BSI regime the ponderomotive energy $U_P = F^2/(2\omega^2)$ (the time-averaged kinetic energy of a classical free charge oscillating in an electromagnetic plane-wave field) of the ionized electron may be greater than its binding energy. This depends both on the laser field parameters and on the principal quantum number $n$ of the initial state. There is a dimensionless parameter $z_1 = 2U_P/E_B$, which measures the ratio of these two energies [25] (this parameter is connected with the Keldysh parameter: $z_1 = 2/\gamma^2$ for CP fields). In principle, one expects that the strong-field approximation (SFA) [14,25,26] (or the Keldysh-Faisal-Reiss model) should work, if $z_1 > 1$ (and better when $z_1 \gg 1$). However, when the SFA (in the velocity gauge) is extended to treat excited bound states of the hydrogen atom one obtains the same total ionization rates and energy distributions of photoelectrons for the initial states $(2,1,-1)$ and $(2,1,1)$. This theoretical result is, of course, unphysical [27]. On the other hand, different ionization rates and energy distributions for these states have been obtained in the length gauge [27,28]. But we have numerically verified that ionization rates for the initial states $(2,1,-1)$ and $(2,1,1)$ in the length gauge are too large [29] to explain any ionization probabilities shown in Figs. 1(a) and 1(b) in the present work and in Figs. 1 and 2 in Ref. [9]. Perhaps the main reason of this drawback of the SFA is the neglect of Coulomb field in the final state of outgoing electron. We are not aware of existing any simple Coulomb correction to the SFA for laser intensities and the laser frequency $\omega = 0.114$ a.u. considered here. For example, when $I = 10^{14} W/cm^2$ one obtains $z_1 = 0.9$, 2.0, and 3.5 for $n = 2$ (Figs. 1(a) and 1(b)), $n = 3$ (Figs. 1(c) and 1(d)), and $n = 4$ (Figs. 1(e) and 1(f)), respectively. Thus, both $U_P$ and $E_B$ have comparable magnitudes, which makes theoretical description of the process of ionization and excitation very difficult.

Coulomb corrections to the SFA were discussed in the recent review of Karnakov *et al.* [30]. Much progress has been achieved in recent years [30,31] (and references therein) particularly for linearly polarized laser field, but for arbitrary elliptical polarization as well. The imaginary time method (which is usually used in calculating probability of the tunneling through a time-dependent barrier) gives results which are considerably different from those based on the SFA [30]. Maybe another factor should be taken into account to explain ionization (and incidentally excitation) probabilities. To compute properly ionization probabilities utilizing the SFA sometimes a modification of the Fermi golden rule or slowly varying population approximation [32,33] is required. In the case of the linear polarization inhibition of ionization by coherent population trapping can occur. One may speculate about analogous effect for the CP light. If the laser pulse is not too short (at least a few cycles) and the peak laser field not too strong, behavior of ionization probability may be understood in terms of the so-called single-state Floquet theory [34,35], which may give results very close to those from numerical solution to the TDSE [20]. Low-lying circular states in the CP light were successfuly described using Floquet methods [34-37]. Such methods help understand the physical mechanisms that determine the photoelectron spectra and the circular dichroism (see Sec. IV) in two-color ionization of the helium ion by CP laser light [38].

Therefore, we will try to explain qualitatively what excited bound states of the hydrogen atom are mainly populated after the end of the laser pulse. There are the following dipole selection rules for the initial circular state $(n,l,m)$ when a single photon (of a positive helicity) is absorbed: $\Delta m = +1$, and $\Delta l = \pm 1$ (if $\Delta m = -1$ the same photon would be emitted by the atom). As noticed in Ref. [39] (cf. Fig. 1 therein) there are two lowest-order pathways for ionization by counter-rotating light and only one lowest-order pathway for corotating light. Higher-order pathways are less probable. Absorbing a few photons may lead to the following final excitations for the initial state with $n = 2$:

$$(l,m) = (1,-1) \to (0,0) \to (1,1) \to (2,2) \to (3,3) \to ... \tag{6}$$

or

$$(l,m) = (1,-1) \to (2,0) \to (3,1) \to (4,2) \to (5,3) \to ... \tag{7}$$

for counter-rotating light and

$$(l,m) = (1,1) \to (2,2) \to (3,3) \to (4,4) \to (5,5) \to ... \tag{8}$$

for corotating light. (The mixing of two paths (6) and (7) is also allowed [39].) Of course, when the laser field is on, the photons may be absorbed or emitted time and again and both quantum numbers $(l,m)$ may decrease as well. For example, the following process is possible: $(l,m) = (1,-1) \to (2,0) \to (1,1) \to (2,2) \to (1,1)$. This is the process of absorption of three photons and then emission of one photon (of the same helicity). This process contributes to net absorption of two photons [40]. Another process, which is possible is the following one: $(l,m) = (1,-1) \to (2,0) \to (1,-1) \to (2,0) \to (1,1)$. This is the process of absorption of one photon, then emission of one photon (of the same helicity) and again two consecutive absorptions of one photon. This process also contributes to net absorption of two photons [40]. Unlike in Ref. [24], we are interested only in net photon absorption after the end of the pulse in this work (without studying populations of bound and excited states when the laser field is on).

### C. Detailed analysis of excitations

In Fig. 1 we present, among others, the excitations as a function of the peak laser intensity. We have looked closer to our numerical data, where populations (after the end of the pulse) of many bound states $(n,l,m)$ with $n \gg 1$ (if necessary) are given. One can draw several conclusions from these data. In general, we have taken a good look to at least three points at each of the six excitation curves from Figs. 1(a)-1(f). One point is located in the maximum (or very close to it), the second one in some perturbative (in relation to excitation) intensity and another one in some large intensity near the right end of each curve. (We have also checked some intermediate points for each curve but we have not found any unexpected behavior.) Generally, in this work we will call the process "$(k = k_0)$ transition" if the atom goes along from the state $(n,l,m)$ to the state $(n',l',m' = m + k_0)$, as a result of interaction with the laser pulse and after its end. We exclude here the case when $n' = n$, $l' = l$, $m' = m$ simultaneously. Of course, if $k_0 > 0$ (hence $m' > m$), it means that $k_0$ photons were net absorbed by the initially excited atom during the laser pulse. If $k_0 < 0$ (hence $m' < m$), it means that $|k_0|$ photons were net emitted. If $k_0 = 0$ (hence $m' = m$), it means that zero photons were net absorbed or emitted (in this case $n' \neq n$ or $l' \neq l$). Usually $n' \geq n$, but $n' < n$ is not excluded, although unlikely. In Tables I-VI, for the above-mentioned six initial states, we present excitation probabilities (i.e., populations in all bound states except the initial one). We also present which of the $(k = k_0)$ transition channels dominate showing their percentage in the respective excitation probability.

For the initial state $(2,1,-1)$ (Table I) $(k=1)$ transition always dominates. The final states are usually $(n,2,0)$ or $(n,0,0)$ for all intensities from Table I. This is in agreement with the pathways (7) and (6), respectively. The range of principal quantum numbers is quite broad, for example, 4, $5 \leq n \leq 16 - 36$ (here and in further text we mention any bound state if its population is at least equal to $10^{-4}$). This is quite anticipated because the difference between the binding energy and the photon energy is small: $E_B - \omega = 0.125 - 0.114 = 0.011$ a.u. and the pulse has some spectral width. For the most intense fields, like $I = 6.0 \times 10^{15} W/cm^2$, the same $(k=1)$ transition still dominates, but the final states of the type $(n,1,-1)$, $(n,1,1)$, and $(n,3,-1)$ also appear in a very visible way. As a result, in strong laser field we have $(k=0)$ and $(k=2)$ transitions as well (and also much weaker $(k=3)$ and $(k=-1)$ transitions; cf. Table I).

For the initial state $(2,1,1)$ (Table II) the situation regarding the excitation is similar to that of the state $(2,1,-1)$, namely $(k=1)$ transitions always dominate for four lower intensities from Table II (leading to $(n,2,2)$ final states with $4 \leq n \leq \sim 30$; this is in agreement with the pathway (8)). But for the highest intensity in Table II, $I = 6.0 \times 10^{15} W/cm^2$ $(k=0)$ transitions are more probable than $(k=1)$ transitions (it means that joint population of $(n,1,1)$ ($n \neq 2$) or $(n,3,1)$ states is greater than the population of $(n,2,2)$ states). There are also weaker $(k=-1)$, $(k=2)$, and $(k=-2)$ transitions.

For the initial states with $n = 3$ or $n = 4$ (Tables III-VI) the situation regarding the excitation is very different from that of the $n = 2$ initial states. We note that if the laser intensity increases one first observes a strong ionization and then the excitation which achieves a local maximum at roughly the same intensity $I \approx 1.0 \times 10^{16} W/cm^2$ (cf. Figs. 1(c)-1(f)). The highest excitations amount to $25 - 40\%$ (depending on $(n,l,m)$), so they are visibly smaller than the excitations for the initial states with $n=2$. When the excitation reaches the local maximum the ionization reaches a local minimum (nearly for the same intensity). But, unlike for the states with $n=2$, the excitation never exceeds the ionization. For each intensity (that we have checked) and each initial state with $n=3$ or $n=4$, $(k=0)$ transitions dominate (cf. Tables III-VI). These excitations are specific by the fact that both numbers $(l,m)$ do not change finally and only $n$ grows after switching-off the laser field. For instance, for the initial state $(4,3,-3)$ and $I=1.0\times 10^{16} W/cm^2$ (very close to maximum excitation for all states with $n=3$ or $n=4$) the population in the states $(n,3,-3)$, where $5\leq n \leq\sim 40$, achieves $0.415$.

There is one more interesting feature of excitations of circular states of the hydrogen atom (Figs. 1(a)-1(f)). This feature is common to three pairs of the initial states which differ by the sign of magnetic quantum number $m$. Namely, atoms in the states with $m<0$ usually absorb more photons (of the positive helicity) than atoms in analogous states with $m>0$ (we have concluded it looking at dominant final values of $m$). This does not mean that the states with $m<0$ have always a bigger probability of excitation. The latter probability depends on intensity for the given principal quantum number $n$. Atoms initially in the states with $m>0$ usually can emit photons (of the same positive helicity), while atoms in analogous states with $m<0$ cannot (at the level of at least $10^{-4}$ of total population). For example, there are almost no $(k<0)$ transitions for the state $(2,1,-1)$, while there are such $(k<0)$ transitions for the state $(2,1,1)$ at $I=6.0\times 10^{15} W/cm^2$ (cf. Tables I and II). There are almost no $(k<0)$ transitions for the state $(3,2,-2)$, while there are such $(k=-1)$ and $(k=-2)$ transitions for the state $(3,2,2)$ (cf. Tables III and IV). There are no $(k<0)$ transitions for the state $(4,3,-3)$, while there are such $(k=-1)$ and $(k=-2)$ transitions for the state $(4,3,3)$ (cf. Tables V and VI).

Thus, the most probable excitation process, for the initial states with $n=2$, is usually $(k=1)$ transition. For higher intensities also $(k=0)$ and $(k=2)$ transitions appear, but in the case of the state $(2,1,1)$ also $(k=-1)$ transition is possible. For the initial states with $n=3$ and $n=4$ usually $(k=0)$ transition takes place and less probable are $(k=1)$ transitions. In the case of the states $(3,2,2)$ and $(4,3,3)$ also $(k=-1)$ transition is possible.

We have also investigated analogically the excitation process for the states $(2,1,\pm 1)$, but for two times lower frequency $\omega=0.057$ a.u. [9] (cf. Figs. 1 and 2 therein; it is interesting to compare those data with the present case of $\omega=0.114$ a.u.). It appears that $(k=2)$ transitions almost always dominate for the initial state $(2,1,-1)$ (with final $(l,m)=(3,1)$ or $(1,1)$), and for the initial state $(2,1,1)$ (with final $(l,m)=(3,3)$). This is in agreement with the above-mentioned pathways (7) or (6) and (8), respectively. Let us note that 2 is the maximum number of energy quanta which can be absorbed by the electron in the state $(2,1,\pm 1)$ without transition to positive energy states ($E_2 + 2\omega = -0.125 + 2 \times 0.057 < 0$; $E_2 + 3\omega > 0$). But for the state $(2,1,1)$ at $I = 2.4 \times 10^{15} W/cm^2$ (which is the highest intensity in Fig. 2 of Ref. [9]) $(k=1)$ transition with final $(l,m)=(2,2)$, is even a little more probable (again in agreement with the pathway (8)). $(k=1)$ transitions are quite probable for both initial states with $n=2$. For higher intensities also $(k=3)$ and $(k=4)$ transitions become probable for the state $(2,1,-1)$ (according to pathways (6) and (7)). On the other hand, for the state $(2,1,1)$ at $I = 2.4 \times 10^{15} W/cm^2$ $(k=0)$, $(k=-1)$, and $(k=-2)$ transitions (with populations of the order of $0.01$) are possible.

### D. Ionization and stabilization

The accompanying process of ionization outweighs the excitation for sufficiently intense fields for $n=2$ and always for $n=3$ or $n=4$. Some insight in stabilization [35,37] of circular Rydberg atoms in CP laser fields was achieved owing to classical trajectory Monte Carlo method and looking at quantum probability density of the atom in the plane perpendicular to the propagation direction of the field [41]. Although the authors of Ref. [41] did their analysis for higher principal quantum numbers, namely $n=5$ and $n=10$, we find their conclusions also useful in our case of low-lying circular states of the atom. In the circular state of the hydrogen atom the radial probability density of the electron has a single maximum located at $r_n = n^2$. There is also a quantum-mechanical average of the electronic distance to the nucleus $\langle r_{nl} \rangle = (3n^2 - l(l+1))/2$, which is a little bigger for circular states and is equal to $n^2 + n/2$. The radial probability density of the electron has a certain significant width around the maximum. In Ref. [41] it was found that a local maximum in the ionization probability occurs if $\alpha_0 \approx \langle r_{nl} \rangle$, where $\alpha_0 = F/\omega^2$ is a quiver radius (of a free electron in the laser field). Very briefly, the ionization is more efficient, when the bound electron is closer to the Coulomb center (the nucleus) as a result of interaction with the laser field. In counter-rotating case ($m < 0$) the electron is dragged outside the torus-shaped initial state of the atom (away from the nucleus), while in corotating case ($m > 0$) the electron is dragged inside the torus (we mean a sphere with the nucleus in the origin), closer to the nucleus. This mechanism is more and more efficient if the quiver radius $\alpha_0$ grows (with a growth of the laser field) from zero to $\langle r_{nl} \rangle$, approximately (cf. Fig. 4 in Ref. [41]). When $\alpha_0 > \langle r_{nl} \rangle$ this mechanism becomes less efficient, because the probablity density in the initial state decreases for $r > \langle r_{nl} \rangle > r_n$.

For $n = 2$, 3, and 4, one obtains $I = 3.0 \times 10^{14} W/cm^2$, $1.3 \times 10^{15} W/cm^2$ and $3.8 \times 10^{15} W/cm^2$, respectively from the equation $\alpha_0 = \langle r_{nl} \rangle$ (and using $I = 2F^2 = 2\alpha_0^2 \omega^4$). Indeed, in Figs. 1(c) and 1(d) we find the local maxima at ionization curves near $I \approx 10^{15} W/cm^2$ which is close to the above prediction for $n = 3$. In Figs. 1(e) and 1(f) there are indistinct local maxima at ionization curves near $I \approx 2 - 4 \times 10^{15} W/cm^2$ which is also close to this prediction for $n = 4$. However, the ionization still grows for higher intensities reaching its greatest values for $I \approx 3 - 4 \times 10^{16} W/cm^2$ (where $\alpha_0 \approx 50 - 58 > \langle r_{43} \rangle = 18$), so maybe some other mechanism works here. Simultaneously, the excitation decreases and we note that for $n = 4$ $I \approx 3.0 \times 10^{15} W/cm^2$ is the intensity at which initial-state probability starts to fall rapidly to zero (cf. Figs. 1(e) and 1(f)). Finally, for $n = 2$ and above $I = 3.0 \times 10^{14} W/cm^2$, the ionization still grows what accompanies decrease of the excitation. The initial-state probability is already very close to zero above $I = 3.0 \times 10^{14} W/cm^2$ (or even for lower intensities).

According to Gavrila [35], quasistationary (or adiabatic) stabilization is connected with the fact that ionization rates decrease with intensity, while dynamic stabilization expresses the fact that ionization probabilities after the end of the laser pulse (of fixed shape and duration) do not approach unity with the growth of the peak intensity. In Figs. 1(a)-1(f) dynamic stabilization is evident in the limited range of laser intensities. According to Popov *et al.* [37], stabilization occurs, if the total probability of ionization per pulse becomes a decreasing function of the peak intensity. Of course, this is true for some intervals of laser intensities in Figs. 1(a)-1(f). There are a lot of papers studying different physical mechanisms of stabilization [35,37] (and references therein) just like interference stabilization of Rydberg atoms and Kramers-Henneberger stabilization or high frequency stabilization of neutral atoms and negative ions. For example, in Ref. [42] Piraux and Potvliege showed that the hydrogen atom, initially in $(5,4,4)$ state, and ionized by 620 $nm$ 120-cycle linearly polarized laser pulse, may be described with high accuracy by numerical solution to the TDSE and by the above-mentioned (time-independent) single-state Floquet theory. Calculations of final ionization probabilities, made in two very different ways, agree in at least three significant digits for peak intensities from perturbative ones up to $I = 2.0 \times 10^{14} W/cm^2$ [42]. In this case only one photon ($\omega = 0.073$ $a.u.$) is needed to ionize the excited hydrogen atom in perturbative regime and excitations to other bound states play no role. Thus, we may suppose that whenever (in our present case of circular polarization) the excitation is negligible, the single-state Floquet method (or maybe its multistate generalization [35]) could be useful. It is well known that an agreement between Floquet method and numerical solution to the TDSE holds also for much shorter pulses. Another kind of stabilization is assumed in the model of interference stabilization [43]. Namely, interference of different-order continuum-continuum and bound-continuum transitions may lead to stabilization. In the paper of Tikhonova *et al.* [43] circular states of the hydrogen atom in linearly polarized laser field were investigated for not too strong fields (when $\alpha_0 <\sim r_n$). The so-called stabilization thresholds (in two particular cases) were calculated in the framework of perturbation theory and good agreement with earlier calculations (one of them was based on numerical solution to the TDSE) was found. The stabilization was explained by interference of different-order transitions to and in the continuum. In both papers [42] and [43] the condition $\omega > E_B$ was satisfied. It seems that both theoretical methods [42,43] could be applied in the present case of circular polarization for the initial states with $n = 3$ or $n = 4$ when total excitation is negligible.

## IV. SUMMARY

A few years ago Herath *et al.* [44] observed experimentally in argon atoms that strong-field ionization rate in the CP laser field depends on the sign of the magnetic quantum number $m$. Later another experimental data [45-47] showed that the ionization probability is dependent on the sign of the magnetic quantum number. Altogether, the different response of a bound system to right- and left-circularly polarized light is called the circular dichroism. Theoretically such effects were predicted much earlier [3,19,20] and quite recently, for instance, in Refs. [9,12,13,24,27,28,38,39,41,48-54]. Most of the above works is devoted to the circular dichroism in adiabatic or nonadiabatic tunneling regimes. Only a few works explored the region that with increasing intensity switches from MPI to BSI. This is also the case of the present paper. In both tunneling regimes excitations of the initial state in the laser field are negligible, while they are very important for most intensities here (if $n = 2$) or at least for higher intensities (if $n = 3$ or $n = 4$).

In conclusion, we have investigated ionization and excitation of the low-lying circular states of the hydrogen atom by the CP 400 *nm* 10-cycle laser pulse. The peak intensities of the pulse have covered about five orders of magnitude, and fixed and positive helicity of the laser light has been used in our calculations. Thus, the $(n,l,m)$ initial states have counter-rotating electrons (with respect to the laser light) if $m < 0$, and corotating electrons if $m > 0$. Our TDSE code has provided us with exact numerical results in the nonrelativistic and dipole approximations. The dominant physical mechanism, which is valid not only in the perturbative regime of intensities, is multiphoton ionization and excitation. We describe and explain our results by showing pathways and final states of the excitation. Almost for all laser field intensities and 400 *nm* radiation the probability of ionization is greater for the state with $m > 0$ than for the state with $m < 0$. Presumably, this is connected with the fact that for $\omega = 0.114$ *a.u.* the condition given by Eq. (1) is very well satisfied (i.e., $\omega \gg \omega_{\lim}$) for the initial states with $n = 2$, 3 and 4. If the ratio $\omega / \omega_{\lim}$ is smaller, like for $\omega = 0.057$ *a.u.*, it becomes possible that the ionization probability for the state $(2,1,-1)$ is visibly greater than that for the state $(2,1,1)$ (c.f. Fig. 5 in Ref. [9]). In what follows, we list the rest of our particular findings.

First, for the initial states $(2,1,-1)$ and $(2,1,1)$ the ionization and the excitation by laser of frequencies $\omega = 0.114$ *a.u.* (400 *nm*) and $\omega = 0.057$ *a.u.* (800 *nm*) are qualitatively similar. With increasing the intensity the ionization starts near the BSI intensity. The excitation starts for about 10 times smaller intensities and may achieve large values ($> 50\%$), larger than the ionization in some range of laser intensities. Eventually, for higher intensities, the ionization prevails. Large excitations are possible mainly owing to $(k = 2)$ transitions for $\omega = 0.057$ *a.u.* and $(k = 1)$ transitions for $\omega = 0.114$ *a.u.* A common thing for these two cases of excitation is the fact that there are a lot of final states which may be populated in agreement with selection rules for absorption of photons and the condition that the final energy should be negative.

Secondly, for the initial states with $n = 3$ or $n = 4$ (and 400 *nm* radiation) the BSI intensity does not play a role because the ionization starts at $I \gg I_{BSI}$. The excitation is much smaller than that for the initial states with $n = 2$ and never exceeds the ionization. With increasing the intensity the excitation starts after the ionization for $n = 3$ or $n = 4$. The main reason is the fact that net absorption of one photon (during the pulse) for $\omega = 0.114$ $a.u.$ usually leads to ionization, but not to excitation if $n = 3$ or $n = 4$. In this case the excitation is dominated by $(k = 0)$ transitions (i.e., initial and final magnetic quantum numbers are the same).

Thirdly, for any initial state considered here $(k = k_0)$ transitions with small $|k_0|$ dominate. The sign of $E_n + k\omega$ ($k \geq 0$ - the number of net photons absorbed) may be positive or negative which usually leads to ionization or excitation, respectively. Regarding only the excitation, based on all particular cases (with almost no exceptions) considered in this work, the following rule may be formulated. The most probable excitation process is $(k = k_0)$ transition where $k_0$ is the largest nonnegative integer such that $E_n + k_0\omega < 0$. If $E_n + \omega > 0$ (so $k_0 = 0$), $(k = 0)$ transition is the most probable channel (this is the above-mentioned case of $n = 3$ or $n = 4$ and $\omega = 0.114$ $a.u.$). For perturbative laser fields, the rule is strict. This rule usually also holds for stronger fields, but with increasing the laser intensity some other channels may appear with comparable probability.

Fourthly, for the circular states with $m > 0$ (corotating electrons) and more intense fields considered in this work it becomes probable that atoms emit net one or two photons (i.e., $m' = m - 1$ or $m' = m - 2$, where $m'$ is the magnetic quantum number of the final bound state).

Fifthly, for the initial circular states with $n = 2$ and intensities $I > \sim 10^{14} W/cm^2$ and for the initial circular states with $n = 3$ or $n = 4$ and intensities $I > \sim 10^{16} W/cm^2$ the initial-state probability is very close to zero (for the laser pulse considered here).

## ACKNOWLEDGMENTS

The authors are indebted to Bernard Piraux for the possibility of utilizing his numerical code in this research. The authors thank both referees for valuable comments and suggestions to improve this work. The present paper has been supported by the University of Lodz.

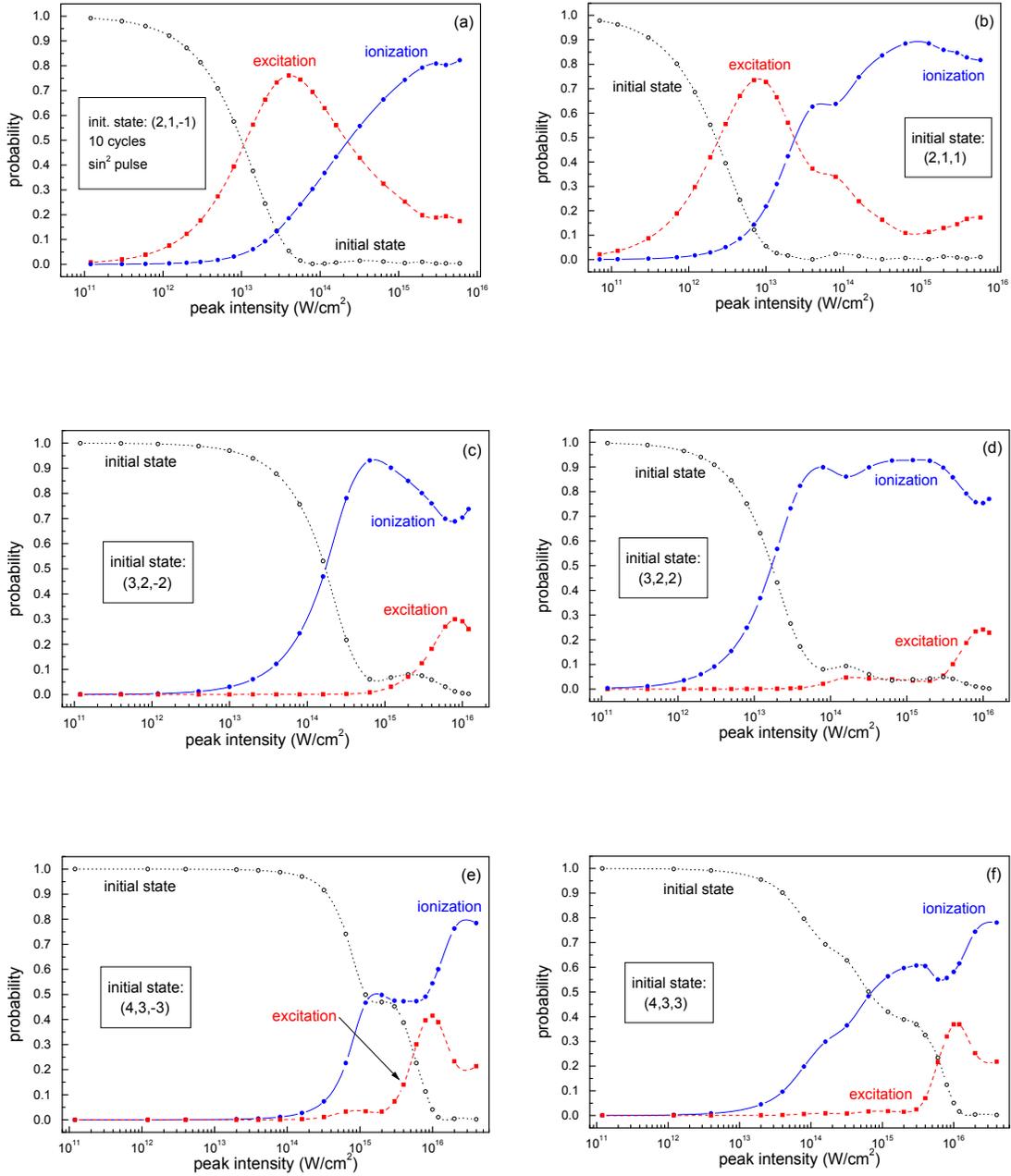

FIG. 1. (Color online) Probability of ionization (blue solid line with solid circles), excitation (red dashed line with solid squares) and remaining in the initial state of the hydrogen atom (black dotted line with open circles) as a function of the peak laser intensity (at the end of the laser pulse). One numerical solution to the TDSE corresponds to each peak intensity on the plot. We have calculated connecting lines with the help of splines. Left panel: counter-rotating light. Right panel: corotating light.

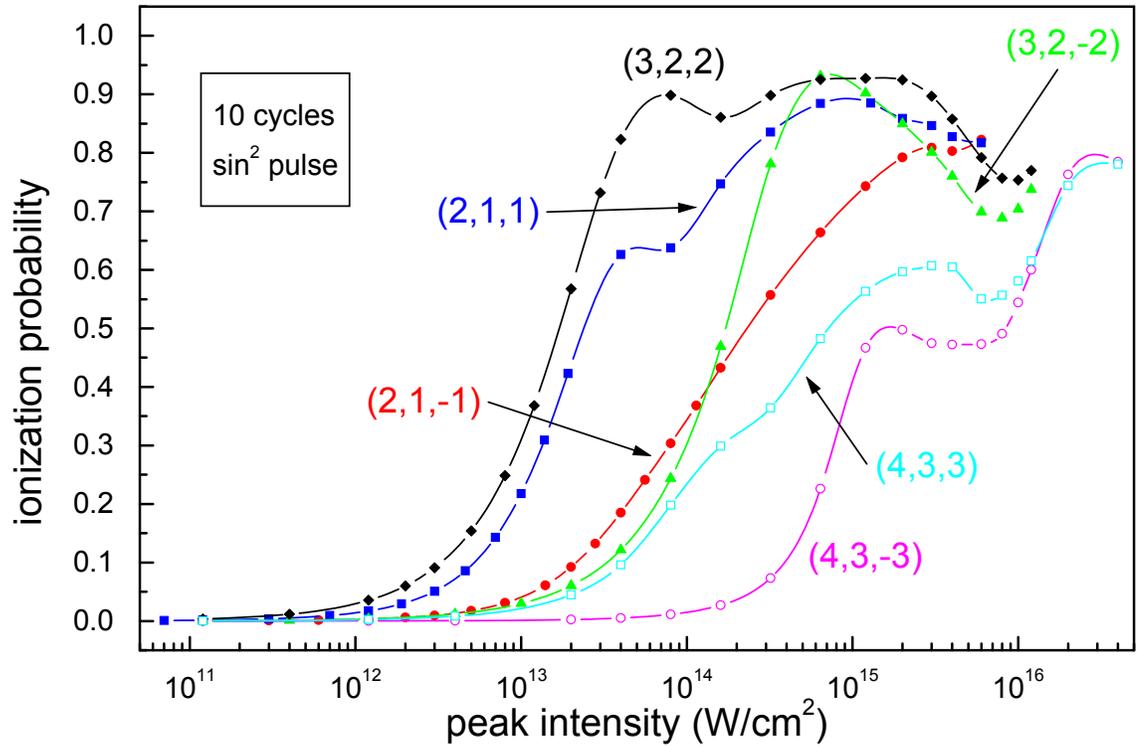

FIG. 2. (Color online) Comparison of ionization probabilities for the six initial states of the hydrogen atom as a function of the peak laser intensity (these lines are identical with respective lines in Fig. 1. $(2,1,-1)$ - red line with solid circles; $(2,1,1)$ - blue line with solid squares; $(3,2,-2)$ - green line with solid triangles; $(3,2,2)$ - black line with solid diamonds; $(4,3,-3)$ - magenta line with open circles; $(4,3,3)$ - cyan line with open squares.

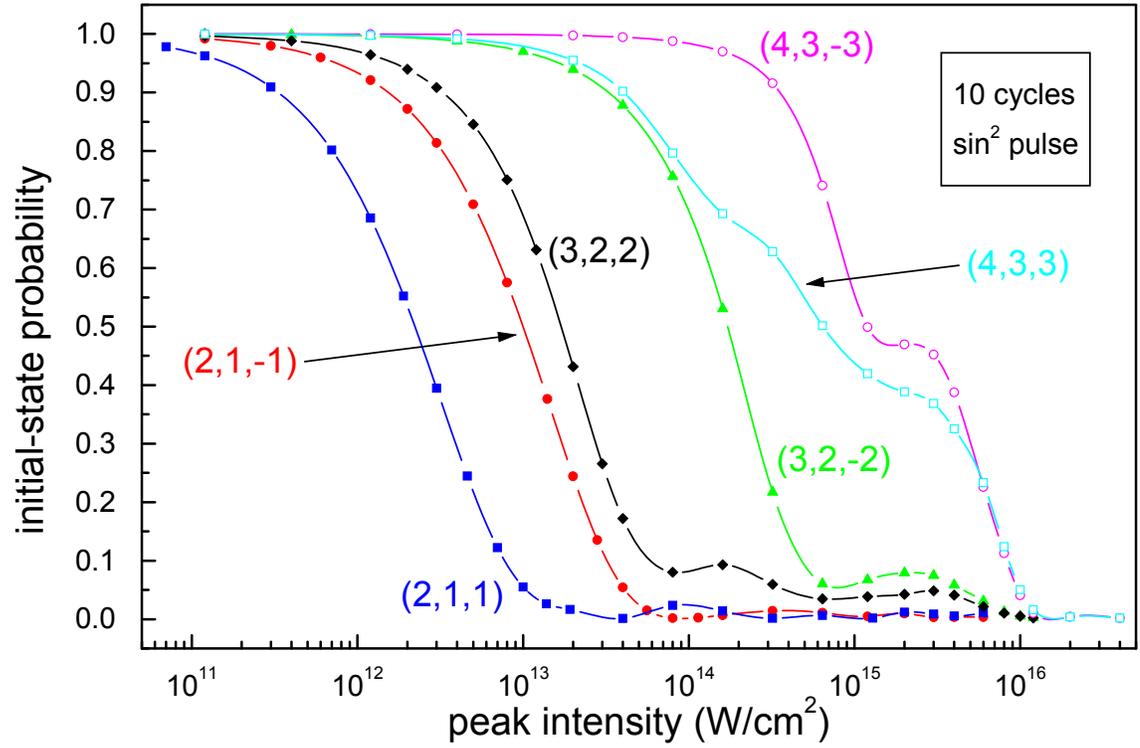

FIG. 3. (Color online) Comparison of initial-state probabilities for the six initial states of the hydrogen atom as a function of the peak laser intensity (these lines are identical with respective lines in Fig. 1). Markers of the lines for the six initial states of the hydrogen atom are the same as in Fig. 2.

TABLE I. The dominant $(k = k_0)$ transition channels for the initial state $(2,1,-1)$ for a few peak intensities (in the first column) of the laser pulse. The total excitation probability (EP) is given in the last column (this is the population in all bound states except the initial one). There are fractions of the total EP for a given intensity in consecutive columns. These fractions should sum up to 100% in a given line (or nearly to 100%, because we discount very unlikely $(k = k_0)$ transitions; we put 0 if the respective fraction is less than 0.01%). For example, 1.2[12] denotes that the peak intensity is $I = 1.2 \times 10^{12} W/cm^2$.

| $I(W/cm^2)$ | $(k=-1)$ | $(k=0)$ | $(k=1)$ | $(k=2)$ | $(k=3)$ | EP |
|---|---|---|---|---|---|---|
| 1.2[12] | 0 | 0 | 100% | 0 | 0 | 0.0754 |
| 8.0[12] | 0 | 0 | 100% | 0 | 0 | 0.394 |
| 4.0[13] | 0 | 0 | 100% | 0 | 0 | 0.761 |
| 3.2[14] | 0 | 0 | 99.99% | 0 | 0 | 0.429 |
| 6.0[15] | 0.03% | 20.2% | 73.4% | 6.0% | 0.35% | 0.174 |

TABLE II. As Table I, but for the initial state $(2,1,1)$.

| $I(W/cm^2)$ | $(k=-2)$ | $(k=-1)$ | $(k=0)$ | $(k=1)$ | $(k=2)$ | EP |
|---|---|---|---|---|---|---|
| 3.0[11] | 0 | 0 | 0 | 100% | 0 | 0.0869 |
| 1.9[12] | 0 | 0 | 0 | 100% | 0 | 0.419 |
| 7.0[12] | 0 | 0 | 0 | 100% | 0 | 0.735 |
| 8.0[13] | 0 | 0 | 0.55% | 99.45% | 0 | 0.339 |
| 6.0[15] | 0.11% | 10.9% | 55.0% | 32.0% | 1.5% | 0.172 |

TABLE III. As Table I, but for the initial state $(3,2,-2)$.

| $I(W/cm^2)$ | $(k=-1)$ | $(k=0)$ | $(k=1)$ | $(k=2)$ | EP |
|---|---|---|---|---|---|
| 2.0[15] | 0.023% | 97.98% | 1.87% | 0.067% | 0.0710 |
| 8.0[15] | 0.014% | 99.00% | 0.92% | 0.058% | 0.299 |
| 1.2[16] | 0.044% | 98.71% | 1.18% | 0.062% | 0.260 |

TABLE IV. As Table I, but for the initial state $(3,2,2)$.

| $I(W/cm^2)$ | $(k=-3)$ | $(k=-2)$ | $(k=-1)$ | $(k=0)$ | $(k=1)$ | $(k=2)$ | EP |
|---|---|---|---|---|---|---|---|
| 6.4[14] | 0 | 0.04% | 19.92% | 67.94% | 12.10% | 0 | 0.0399 |
| 6.0[15] | 0.074% | 1.88% | 10.22% | 85.08% | 2.74% | 0 | 0.187 |
| 1.2[16] | 0.046% | 2.14% | 8.32% | 85.75% | 3.74% | 0 | 0.228 |

TABLE V. As Table I, but for the initial state $(4,3,-3)$.

| $I(W/cm^2)$ | $(k=-1)$ | $(k=0)$ | $(k=1)$ | $(k=2)$ | EP |
|---|---|---|---|---|---|
| 3.0[15] | 0 | 99.79% | 0.20% | 0 | 0.0732 |
| 1.0[16] | 0 | 99.93% | 0.065% | 0 | 0.415 |
| 4.0[16] | 0 | 99.66% | 0.34% | 0 | 0.214 |

TABLE VI. As Table I, but for the initial state $(4,3,3)$.

| $I(W/cm^2)$ | $(k=-2)$ | $(k=-1)$ | $(k=0)$ | $(k=1)$ | EP |
|---|---|---|---|---|---|
| 4.0[15] | 0.40% | 1.78% | 97.78% | 0 | 0.0697 |
| 1.0[16] | 0.07% | 0.59% | 99.33% | 0 | 0.369 |
| 4.0[16] | 0 | 0.48% | 99.47% | 0 | 0.217 |